\documentclass[aip, rse, amsmath, amssymb, reprint, trackchanges, usenames, dvipsnames]{revtex4-2}

\makeatletter
\def\@email#1#2{%
 \endgroup
 \patchcmd{\titleblock@produce}
  {\frontmatter@RRAPformat}
  {\frontmatter@RRAPformat{\produce@RRAP{*#1\href{mailto:#2}{#2}}}\frontmatter@RRAPformat}
  {}{}
}%
\makeatother

\usepackage{graphicx}
\usepackage{dcolumn}
\usepackage[utf8]{inputenc}
\usepackage{mathptmx}

\usepackage{etoolbox}

\usepackage{tabularx} 
\usepackage[version=4]{mhchem} 
\usepackage{pifont}
\usepackage{lipsum}
\usepackage{comment}
\usepackage{multirow}
\usepackage{orcidlink}
\usepackage{hyperref}
\hypersetup{
    colorlinks=true,
    linkcolor=blue,
    filecolor=blue,      
    citecolor=blue,
    urlcolor=blue,
}
\makeatletter
\protected\def\xvcenter{%
  \hbox\bgroup$\everyvbox{\everyvbox{}\aftergroup\m@th\aftergroup$\aftergroup\egroup}%
  \vcenter
}
\DeclareRobustCommand{\midscript}[1]{
  \mathchoice{\mid@script\scriptstyle{#1}}
    {\mid@script\scriptstyle{#1}}
    {\mid@script\scriptscriptstyle{#1}}
    {\mid@script\scriptscriptstyle{#1}}
}
\newcommand{\mid@script}[2]{
  \vcenter{\hbox{$\m@th#1#2$}}
}
\DeclareRobustCommand{\textmidscript}[1]{%
  \xvcenter{\hbox{\scriptsize#1}}%
}
\makeatletter

\newcount\my@repeat@count
\newcommand{\blocks}[1]{%
  \textmidscript{
  \begingroup
  \my@repeat@count=\z@
  \@whilenum\my@repeat@count<#1\do{\ding{121}\advance\my@repeat@count\@ne}%
  \endgroup
  \begingroup
  \my@repeat@count=\z@
  \@whilenum\my@repeat@count< \the\numexpr 4 - #1 \do{\:\advance\my@repeat@count\@ne}%
  \endgroup
  }
}
\makeatother

\begin{document}

\title{Transforming Materials Discovery for Artificial Photosynthesis: High-Throughput Screening of Earth-Abundant Semiconductors}

\author{Sean M. Stafford\orcidlink{0000-0001-8010-4202}}
\affiliation{Department of Chemical Engineering and Materials Science, Michigan State University, East Lansing, MI, 48824, USA}

\author{Alexander Aduenko\orcidlink{0000-0002-0959-2052}}

\affiliation{Moscow Institute of Physics and Technology, Moscow, Russia}

\author{Marcus Djokic\orcidlink{0000-0001-6504-9903}}
\affiliation{Department of Chemical Engineering and Materials Science, Michigan State University, East Lansing, MI, 48824, USA}

\author{Yu-Hsiu Lin\orcidlink{0000-0002-3599-9032}}
\affiliation{Department of Chemical Engineering and Materials Science, Michigan State University, East Lansing, MI, 48824, USA}

\author{Jose L. Mendoza-Cortes\orcidlink{0000-0001-5184-1406}}
\affiliation{Department of Chemical Engineering and Materials Science, Michigan State University, East Lansing, MI, 48824, USA}

\email{jmendoza@msu.edu}

\begin{abstract}
We present a highly efficient workflow for designing semiconductor structures with specific physical properties, which can be utilized for a range of applications, including photocatalytic water splitting. Our algorithm generates candidate structures composed of earth-abundant elements that exhibit optimal light-trapping, high efficiency in \ce{H2} and/or \ce{O2} production, and resistance to reduction and oxidation in aqueous media. To achieve this, we use an ionic translation model trained on the Inorganic Crystal Structure Database (ICSD) to predict over thirty thousand undiscovered semiconductor compositions. These predictions are then screened for redox stability under Hydrogen Evolution Reaction (HER) or Oxygen Evolution Reaction (OER) conditions before generating thermodynamically stable crystal structures and calculating accurate band gap values for the compounds. Our approach results in the identification of dozens of promising semiconductor candidates with ideal properties for artificial photosynthesis, offering a significant advancement toward the conversion of sunlight into chemical fuels.

\end{abstract}

\maketitle

\section{Introduction}

Alarmingly, humanity's consumption of fossil fuels continues to grow rapidly despite widespread awareness of their connection to the climate crisis. \citep{ VaclavEnergyConsumption, EnergyConsumption,BPEnergyConsumption} The sun offers the best path to wean ourselves off these pollutants as it provides about as much energy to Earth every hour that humanity uses throughout an entire year. \citep{EnergyConsumption,BPEnergyConsumption,NASAsunFacts} Solar currently remains a discouraging 1.5\% share of our energy consumption, but thanks to investment in the past decade, this share is growing exponentially. \citep{EnergyConsumption,BPEnergyConsumption, VaclavEnergyConsumption}

The vast majority of investment in solar energy has been dedicated to the research and production of photovoltaic (PV) cells, primarily in the form of solar panels. As a result of this investment, the technology has matured significantly and become increasingly accessible. In fact, the price of solar panels has plummeted by over 99.6\% since 1976, when their power generation capacity was a million times less than it is today. This data is supported by multiple sources, including solar panel price and uptake data.\citep{solarpanelspriceuptakeOWID, SourceOfOWID1, SourceOfOWID2, SourceOfOWID3}

\begin{figure*}
    \centering
    \includegraphics[width=1\textwidth]{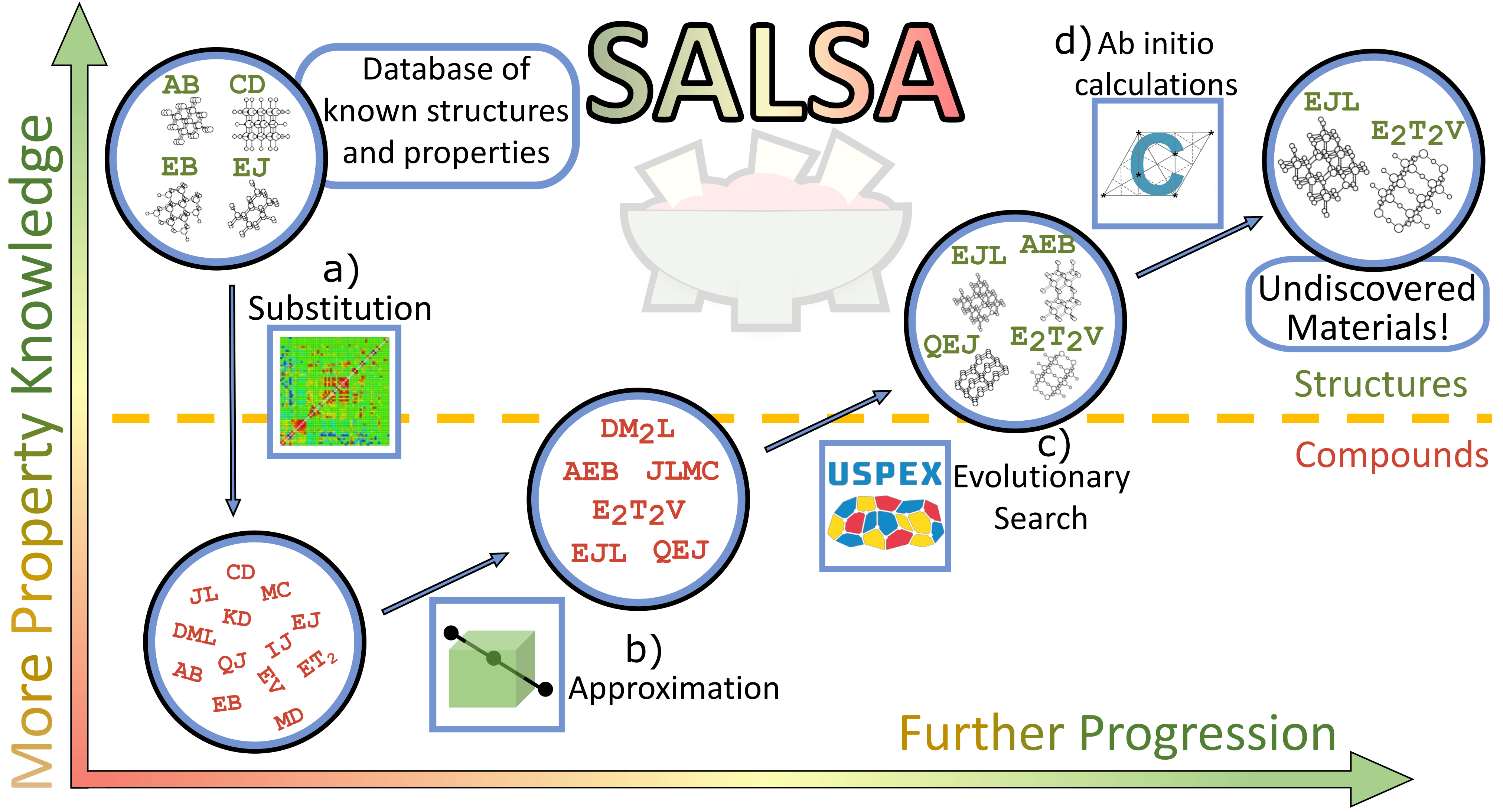}
    \caption{Introducing the SALSA workflow: A Comprehensive Approach to Materials Discovery. Our novel workflow begins with a curated dataset of compounds with known structures and properties. Leveraging an enhanced substitution matrix we constructed from the full ICSD, we generate a vast library of candidate compounds. We then filter these candidates by identifying structural interpolations with desired properties, ultimately using the USPEX algorithm to determine their structures. Lastly, we employ the high-fidelity CRYSTAL software to perform accurate calculations of both structures and properties}
    \label{fig:SALSAoverview}
\end{figure*}

Photovoltaic (PV) cells, while a promising source of renewable energy, face a significant challenge due to their inherent intermittency.\citep{SolarGeneral2018, SolarGeneral2019HRES, SolarGeneral2019SP, SolarGeneral2022} As they generate electricity by converting sunlight into a potential difference between photoelectrode components,\citep{indirectbandgap} they do not store energy, resulting in an output that is dependent on sunlight availability. The power output of PV cells is, therefore, subject to daily and annual oscillations, as well as fluctuations in weather conditions and regional climate differences.\citep{SolarGeneral2018, SolarGeneral2019HRES, SolarGeneral2019SP, SolarGeneral2022}

A promising alternative to traditional solar technology is the photo-electrolyzer. This cutting-edge system harnesses electricity generated by a PV material to power a water-splitting reaction on a catalyst. By separating the functions of trapping sunlight and generating fuel into two distinct components, the photo-electrolyzer generates Hydrogen and Oxygen fuel from sunlight indirectly. This innovative approach circumvents the intermittency problem associated with conventional solar power systems, ensuring energy remains available even when sunlight is not. However, there are still a few hurdles to overcome. For instance, the current system requires wired connections, which can result in significant energy loss. Additionally, the high cost of the water-splitting catalyst (typically made of Platinum or other rare-earth elements) has been a significant barrier to the scalability of photo-electrolyzer technology. A third, unrealized technology - a ``no-wires" photo-electrolyzer system that performs photovoltaic and catalytic functions in a single material - shows great promise. With a cost-effective material, this groundbreaking photocatalytic water-splitting process could address the efficiency and scalability problems of photo-electrolyzers, as well as the intermittency problem of PV cells.

This paper outlines our quest for a breakthrough photocatalytic water-splitting material that meets the critical requirements of stability, efficiency, and scalability. Unfortunately, no existing material is currently able to meet all these essential criteria. Our search is guided by the demanding specifications of the artificial photosynthesis process we are striving to achieve. To effectively split water, a photocatalyst must possess discrete electronic excitations, which require a semiconductor material. The material's electronic structure governs photoabsorption, with the band gap $E_g$ acting as a filter for lower energy photons that are unable to promote an electron to the conduction band and initiate an excitation. To achieve maximum photoabsorption rates, an efficient photocatalyst must be sensitive to light in the high solar availability range of approximately 1-3 eV. Furthermore, the band gap must be direct to ensure optimal performance. \citep{PhotoChemistry, PhotolysisWaterSeminal, indirectbandgap}
In addition to electronic properties, the material must also exhibit excellent stability in an aqueous solution. The photocathode may undergo a reduction reaction with itself and decompose if its reduction potential $\phi_{red}$ is positive relative to the Normal Hydrogen Electrode (NHE). Similarly, the photoanode may decompose if its oxidation potential $\phi_{ox}$ is less than 1.23 V wrt. NHE, which is the oxidation potential of water. Consequently, the redox potentials of the material must be compatible with aqueous stability requirements.
Finally, any successful artificial photosynthesis technology must be composed of Earth-abundant elements to keep the material cost-effective and accessible. This critical constraint ensures that the material is far cheaper than Platinum, making it more widely available for research and development. \citep{PhotoChemistry}
In summary, our search for the ideal photocatalytic water-splitting material is restricted to Earth-abundant elements that possess compatible redox potentials and band gaps for both aqueous stability and efficient photocatalysis.

In the past, searching for a material with a specific set of properties relied heavily on heuristic models, which often proved inadequate due to the vastness of structure space and the complexity of structure-property relationships. This made the search for an optimal material a daunting task. However, recent advancements in computational techniques, such as the use of modern processing power and sophisticated simulation software, have significantly improved the ability to search structure space more effectively. \citep{ChemistvsMachine} This materials design revolution can be largely attributed to the substantial improvements in density functional theory (DFT), which can now predict the properties of previously unknown materials with reasonable reliability. 
Despite recent improvements in density functional theory (DFT), a brute-force approach to materials discovery remains impractical. However, researchers have developed strategic improvements over brute force methods, such as the use of large databases of known materials to identify patterns and make inferences about new materials to guide the search. \citep{StructureDesignMatDisc,MaterialsDiscoveryML} One such tool in this vein is the substitution likelihood matrix. It was introduced by \citet{SubstitutionMatrix} about a decade ago to assess the plausibility of the existence of compounds that differ from known compounds by the swap of ionic components.  Recently, this tool has been enhanced and updated by Stafford et al. (2023b, in preparation).

Another strategic improvement is the use of structure prediction algorithms, which can significantly improve the efficiency of materials discovery. One such algorithm is the Universal Structure Predictor: Evolutionary Xtallography (USPEX), an evolutionary structure search algorithm that interfaces with a DFT code to generate stable crystal structures for a given composition.\citep{USPEX1, USPEX2, USPEX3} By utilizing structure prediction algorithms like USPEX alongside other strategies and tools, such as large databases of known materials and substitution likelihood matrices, we have designed a novel and more efficient materials discovery process.

This paper aims to not only introduce our novel materials discovery process but also to showcase its practical application in the field of artificial photosynthesis. In Section~\ref{sect:SALSA}, we present SALSA, our systematic approach to materials discovery that combines database mining, substitution likelihood analysis, and evolutionary structure prediction algorithms.
In Section~\ref{sect:SALSAapplied}, we demonstrate the efficacy of SALSA by applying it to the search for a photocatalytic water-splitter, a crucial component of artificial photosynthesis.
In Section~\ref{sect:Discussion}, we analyze and contextualize the results of our application, highlighting the benefits of our approach compared to traditional methods.
Furthermore, in Section~\ref{sect:Methods}, we provide more detailed descriptions of the computational techniques used in SALSA, including density functional theory and crystal structure prediction algorithms. Finally, in Section~\ref{sect:Conclusions}, we conclude with some reflections on the potential impact of SALSA on the development of materials for photocatalytic water-splitting and other important applications in materials science.

\section{SALSA -- (S)ubstitution, (A)pproximation, evo(L)utionary (S)earch, and (A)b-initio calculations} \label{sect:SALSA}

We developed a highly efficient and versatile materials discovery process, dubbed SALSA, which is an acronym for \textbf{S}ubstitution, \textbf{A}pproximation, evo\textbf{L}utionary \textbf{S}earch, and \textbf{A}b-initio calculations. An overview of SALSA is provided in Figure~\ref{fig:SALSAoverview}. The process starts by taking a target property or set of properties as input and returns a set of candidate structures as output. Instead of relying on brute-force approaches, SALSA harnesses the power of a large database of compounds with known structures and properties to rapidly search for new materials. The process begins with swapping ionic components between pairs of known compounds that have similar ionic species, as guided by a substitution likelihood matrix, to produce a dataset of hybrid compounds with defined compositions but undefined structures. We then infer approximate properties for these hybrid compounds using a weighted sum of properties of parent compounds and discard hybrids without desirable properties. Promising hybrids are then subjected to an evolutionary structure search using the USPEX algorithm, which generates stable crystal structures for a given composition whenever possible. High-fidelity DFT calculations are then used to recalculate the properties of the generated structures, and structures with undesirable properties are discarded. The process produces a set of undiscovered materials that are promising candidates for various applications, including the application to artificial photosynthesis discussed in Section~\ref{sect:SALSAapplied}. Furthermore, SALSA is highly versatile and can be applied to other materials science problems as well.

\paragraph{Substitution by Chemical Similarity}

Our group reconstructed and expanded the scope of the substitution likelihood matrix introduced by \citet{SubstitutionMatrix} In our construction, we used the entirety of the Inorganic Crystal Structure Database (ICSD) \citep{ICSD} and do not restrict substitutions to preserve the space group of the crystal structure
(Stafford et al., 2023b in prep will describe details of this construction.)
High values of our matrix correspond to pairs of ionic species empirically observed to exist in similar chemical environments. Above a chosen threshold, a value designates substitution between an ion pair as likely. Applying these likely substitutions to compounds of our initial dataset forms a hypothetical set of new candidate compounds. The resulting candidate dataset is too large for us to feasibly calculate properties of all compounds unless we are overly restrictive with unit cell size or substitution threshold. Therefore, we narrow the scope of our investigation to a subset for which we can efficiently approximate properties.

\paragraph{Approximation by Linear Interpolation} \label{sect:SALSAmethodInterp}

\begin{figure}
    \centering
\includegraphics[width=1\columnwidth]{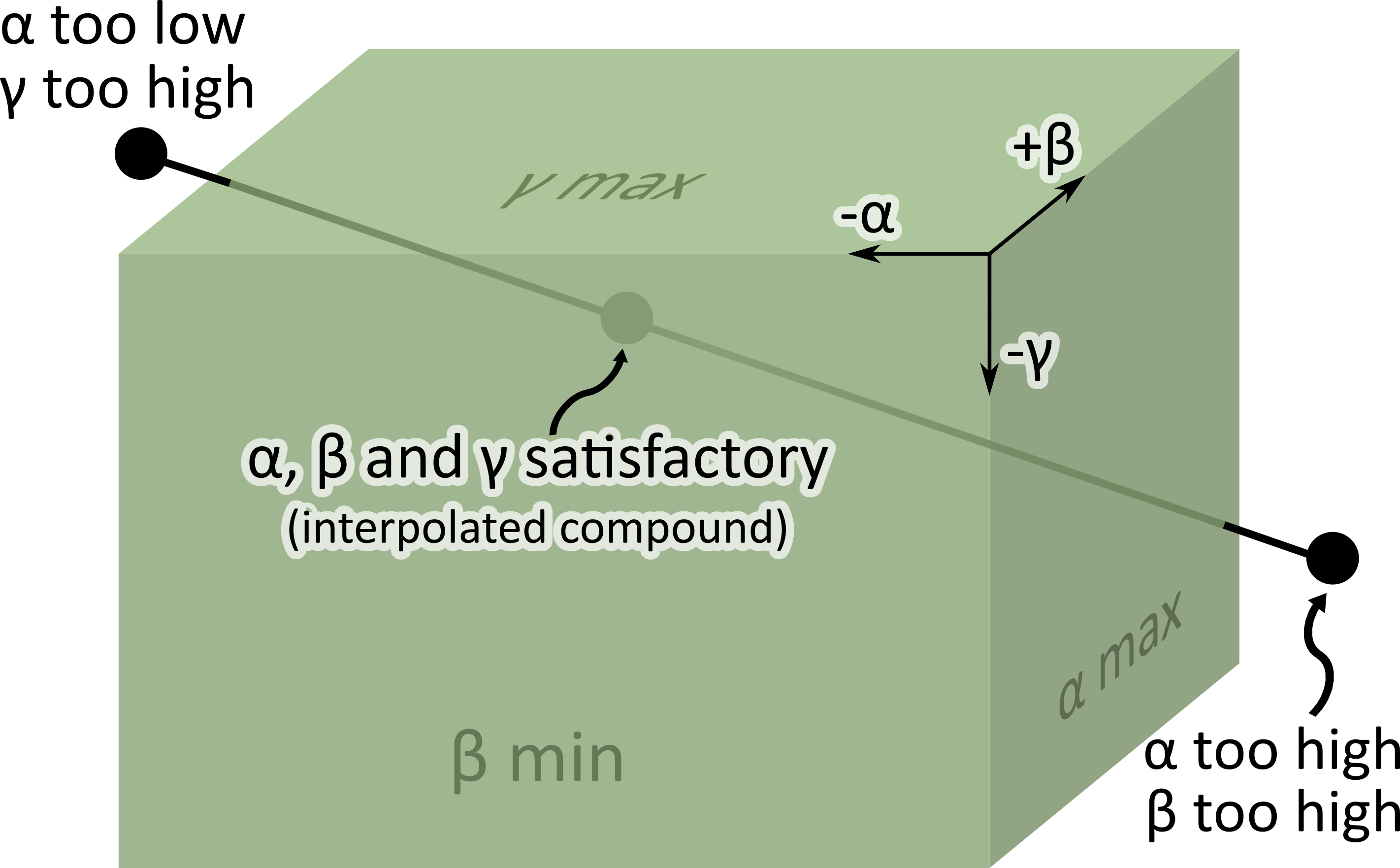}
    \caption{SALSA's composition-property interpolation scheme illustrated for generic properties $\alpha$, $\beta$ and $\gamma$. Parent and hybrid compounds are represented by points outside and within a target region, respectively. Target region represented by green cuboid. For simplicity in depiction, each property has an upper and lower bound here, but this is not required. }
    \label{fig:GenericInterpol}
\end{figure}

We examine the class of candidate compounds which are compositional interpolations between two initial compounds, i.e. hybrid compounds. We derive estimates for the properties of hybrids by summing the properties of parent compounds with the same ratio used in the corresponding hybrid composition. Next, we define the boundary of a target region of property space appropriate for our application. Finally, we eliminate hybrids that do not lie within this region. This step allows us to filter out the sizeable portion of our candidate compounds that are far removed from the target region before proceeding to intensive calculations. While this is an extremely simplistic model of property space, it is a computationally cheap way to approximate values close enough to eliminate most of the unsuitable candidates without a high risk of eliminating suitable ones. Note that we reduce this risk by extending the boundary of our target region beyond the ideal region of property space by enough to include some tolerance for the error that comes with our interpolation method. See Figure~\ref{fig:GenericInterpol} for a summary of this scheme.

\paragraph{Evolutionary Search of Structure Space} \label{sect:SALSA_USPEX}

Until this point, we have defined our hybrid compounds by their composition alone, but reliable property calculations require structural information. Crystal structure prediction from first principles is prohibitively difficult using just composition. 
Instead, we turn to an evolutionary structure search code, USPEX, to generate crystal structures for our hybrids. We provide USPEX with a hybrid composition and enable all available stochastic variation operations, which includes variation of the space group. If USPEX is unable to converge a structure for a given composition, that indicates the composition is unlikely to have a thermodynamically stable structure and is eliminated from further consideration. See Section~\ref{sect:USPEXSettings} for a more detailed look at our USPEX methodology.

\newpage
\onecolumngrid

\begin{figure}[h]
    \centering
\includegraphics[width=.993\columnwidth]{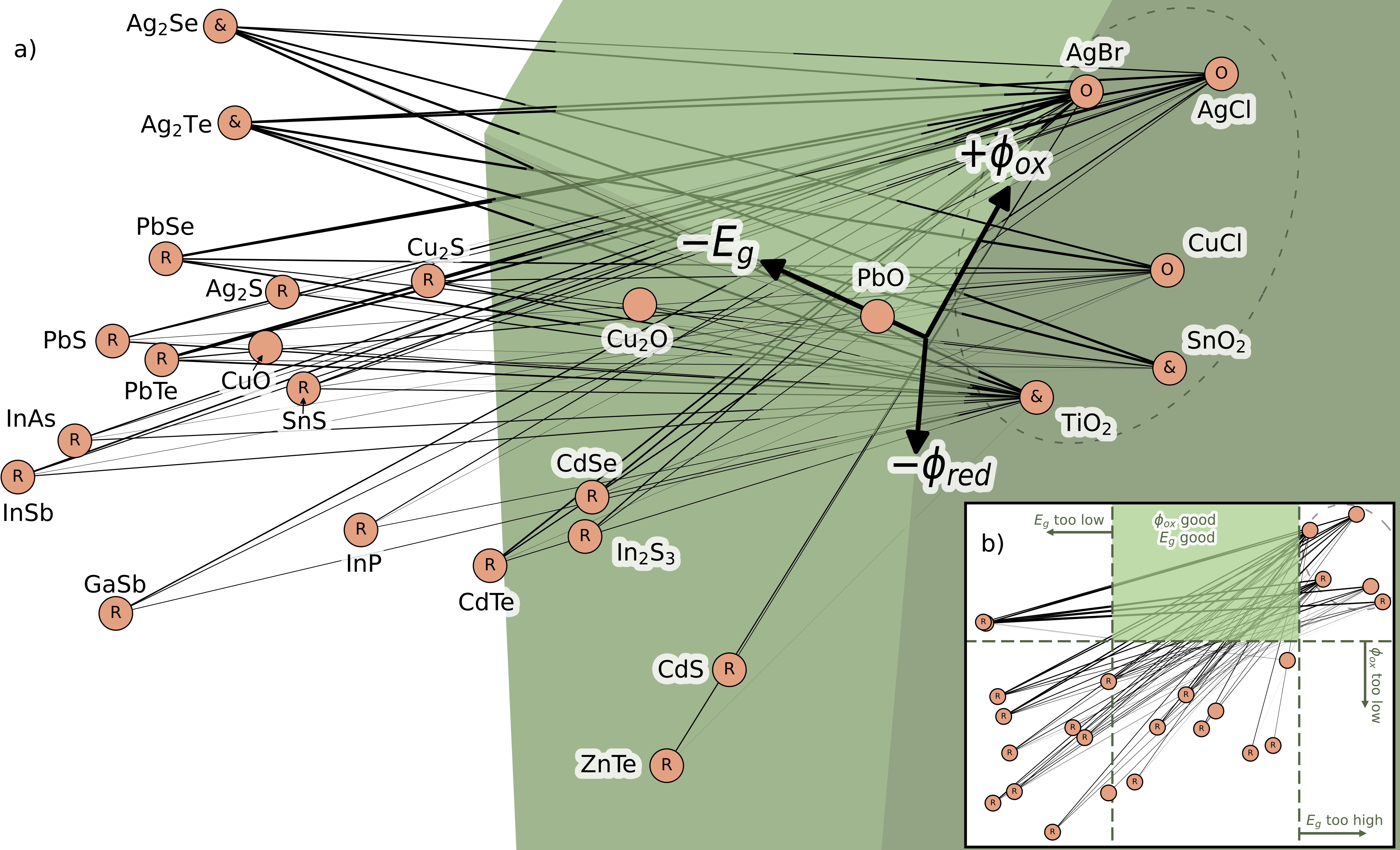}
    \caption{a) A visualization of band gap - oxidation potential - reduction potential space from a perspective that highlights possible interpolations into the ideal property space. Any compound in our initial dataset that could produce one or more interpolations of interest is represented here. Those which had suitable $\phi_{ox}$, $\phi_{red}$  or both are labeled with ``O", ``R" and ``\&", respectively. Lines represent interpolations, with line thickness proportional to a distance within the ideal region. Dashed oval identifies an influential high-$\phi_{ox}$ cluster. Extra 0.2 eV/V boundary region not depicted here. b) A ``top-down" 2D projection of this space excluding the  $\phi_{red}$  dimension.  ``R" indicates a compound with suitable  $\phi_{red}$.   }
    \label{fig:InterpolationDesirablePropertySpace}
\end{figure}

\twocolumngrid

\paragraph{Ab-initio Property Calculations} \label{sect:SALSA_CRYSTAL}
Our candidate set is now vastly narrowed down and contains structural information so 
high fidelity property calculations are computationally feasible. 
Therefore we perform geometry optimization and property calculation with another DFT code, CRYSTAL17, at the hybrid functional level of theory.\citep{CRYSTAL17, CRYSTAL17_2} 
Some candidate compounds located within the target region according to interpolation-inferred values shift outside the region upon replacement by CRYSTAL17-calculated values while others do not converge with CRYSTAL17 at all. We discard these and are left with the final products of SALSA -- the structures which CRYSTAL17 converges and determines to have properties in the target region.

\section{SALSA Applied to Photocatalytic Water-Splitting} \label{sect:SALSAapplied}
We found that millions of candidate compounds could be generated from our initial dataset with the ion exchanges suggested by our substitution matrix. Of these, about 13,600 were compatible with our structural interpolation scheme, that is, they could be constructed as hybrids of compounds within our initial dataset of known semiconductors. See Section~\ref{sect:CandidateParameters} for details on this dataset construction.

\begin{table}[h]
\centering
\caption{A selection of ternary hybrid compounds including silver telluride-bromides, silver sulfide-bromides, titanium cuprates and titanium-lead oxides. All interpolated band gaps and redox potentials lie within target ranges. One \textmidscript{\ding{121}}-symbol appears next to a value for each 0.05 eV/V it lies outside of the ideal range (rounded down).} \label{tab:SubMatrix_Compounds}
\setlength{\tabcolsep}{6pt}
\begin{tabularx}{0.49\textwidth}{lccc}
\hline \hline
Compound         & Band gap  (eV) & Oxidation (V) & Reduction (V) \\  \hline
\ce{Ag2Te} - \ce{AgBr} & & & \\ \hline
\ce{Ag3TeBr}  & \blocks{0} 1.26 \blocks{0}  & \blocks{0} 1.69 \blocks{0}    & \blocks{0} $-$0.41 \blocks{0}   \\
\ce{Ag4TeBr2} & \blocks{0} 1.72 \blocks{0}  & \blocks{0} 1.83 \blocks{0}    & \blocks{0} $-$0.27 \blocks{0}   \\
\ce{Ag5TeBr3} & \blocks{0} 1.98 \blocks{0}  & \blocks{0} 1.90 \blocks{0}    & \blocks{0} $-$0.20 \blocks{0}   \\ \hline
\ce{Ag2S} - \ce{AgBr} & & & \\ \hline
\ce{Ag5SBr3}  & \blocks{0} 2.23 \blocks{0}  & \blocks{0} 1.61 \blocks{0}    & \blocks{0}  \ \ \  0.03 \blocks{1}    \\
\ce{Ag3SBr}   & \blocks{0} 1.70 \blocks{0}  & \blocks{0} 1.17 \blocks{2}    & \blocks{0} $-$0.01 \blocks{0}   \\
\ce{Ag4SBr2}  & \blocks{0} 2.04 \blocks{0}  & \blocks{0} 1.45 \blocks{0}    & \blocks{0}  \ \ \  0.01 \blocks{1}    \\ \hline
\ce{TiO2} - \ce{CuO} & & & \\ \hline
\ce{Ti2CuO5}  & \blocks{0} 2.55 \blocks{0}  & \blocks{0} 1.30 \blocks{0}    & \blocks{0} $-$0.48 \blocks{0}   \\
\ce{Ti3CuO7}  & \blocks{0} 2.67 \blocks{0}  & \blocks{0} 1.42 \blocks{0}    & \blocks{0} $-$0.58 \blocks{0}   \\
\ce{TiCuO3}   & \blocks{0} 2.28 \blocks{0}  & \blocks{0} 1.03 \blocks{4}    & \blocks{0} $-$0.28 \blocks{0}   \\ \hline
\ce{TiO2} - \ce{PbO} & & & \\ \hline
\ce{Ti2PbO5}  & \blocks{0} 2.93 \blocks{3}  & \blocks{0} 1.58 \blocks{0}    & \blocks{0} $-$0.56 \blocks{0}   \\
\ce{TiPb2O4}  & \blocks{0} 2.83 \blocks{1}  & \blocks{0} 1.36 \blocks{0}    & \blocks{0} $-$0.22 \blocks{0}   \\
\ce{TiPbO3}   & \blocks{0} 2.88 \blocks{2}  & \blocks{0} 1.48 \blocks{0}    & \blocks{0} $-$0.40 \blocks{0} \\
\hline\hline
\end{tabularx}
\end{table}

\clearpage
\newpage

\onecolumngrid

\begin{figure}[h]
    \centering
\includegraphics[width=1\textwidth]{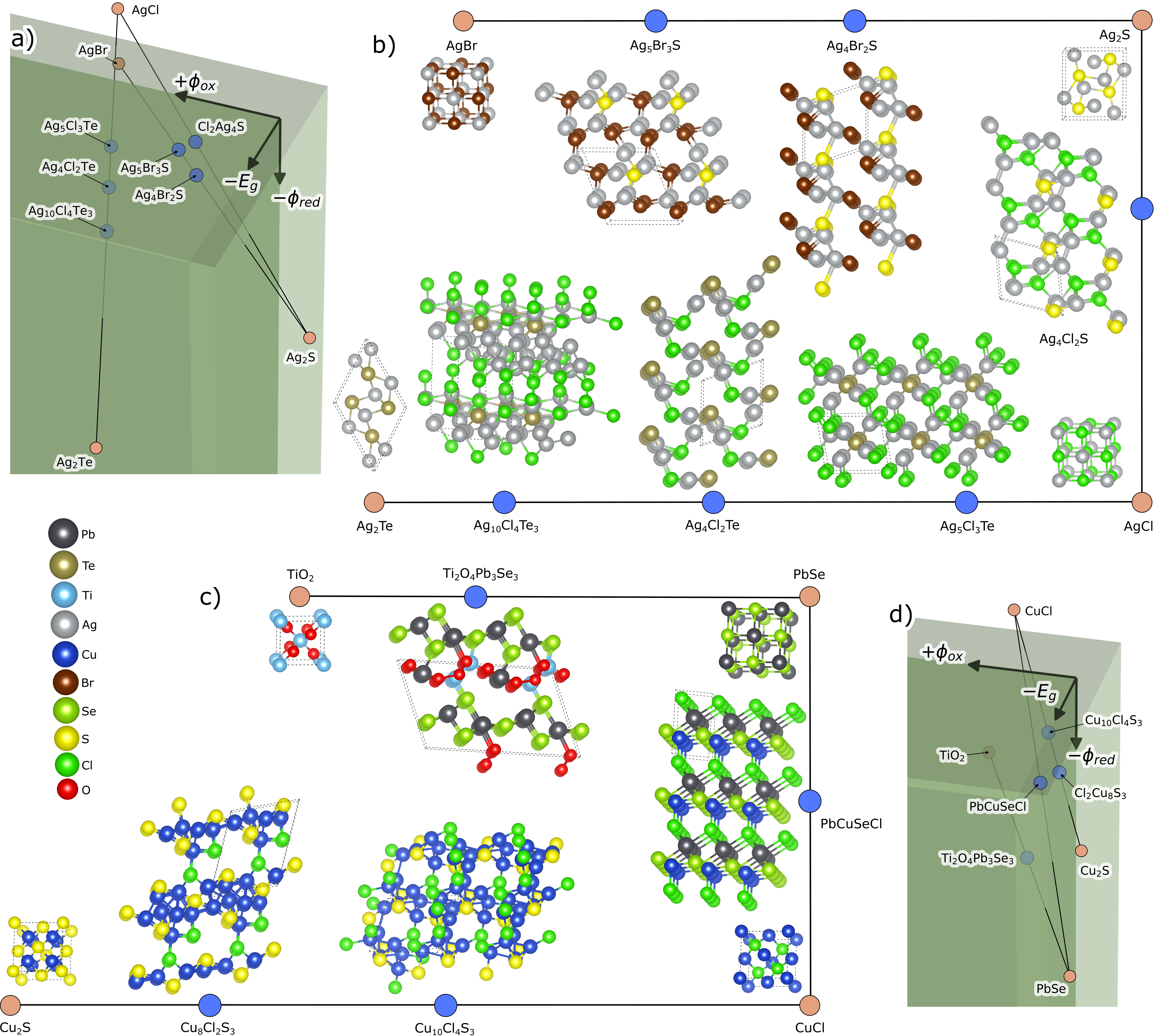}
    \caption{a) and d) A visualization of band gap - oxidation potential - reduction potential space from a perspective which highlights some interpolations that yielded USPEX-converged structures. Depicted hybrid compounds are represented by blue points. Initial compounds that were parents to the depicted hybrid compounds are represented by peach-colored points. Extra 0.2 eV/V boundary region depicted in translucent green.   b) and c) Crystal structures we found for the hybrid compounds. Atom sizes are consistent throughout figure for a given element, except atoms in the legend, which are 2 times as large in diameter. For each structure, dashed gray lines indicate the extent of a single conventional unit cell.}
    \label{fig:USPEX15structures}
\end{figure}

\twocolumngrid

\subsection{Candidate Compounds}\label{sect:SubMatrixResults}

Overall, we found about 1250 hybrid compounds within our target region, including 484 within our ideal region. This corresponds to roughly one out of every 10 and 30 of all possible hybrids, respectively. Most interpolation pairings involved binary compounds with no elements in common so more hybrids were quaternary rather than ternary.  Furthermore, the binary parents of ternary compounds tended to be located more closely to each other in property space, without any portion of the target region between them, so ternary compounds were relatively underrepresented in the regions of interest. The quaternary:ternary ratio was about 5:1 overall, 7:1 in the target region, and 8:1 in the ideal region.


\begin{table*}[t]
\centering
\caption{Sample of compounds converged by USPEX with their interpolated band gaps and redox potentials. One \textmidscript{\ding{121}}-symbol appears next to a value for each 0.05~eV/V it lies outside of the ideal range (rounded down).} \label{tab:USPEXStructs}
\setlength{\tabcolsep}{24pt}
\begin{tabularx}{1\textwidth}{lcccc}
\hline \hline
Compound          & Band gap  (eV) & Oxidation (V) & Reduction (V) & Interpolation                    \\ \hline
\ce{Hybrids} & & & & \\ \hline
\ce{Ag10Cl4Te3}  & 1.63     & 1.81  \blocks{0}      & \blocks{0}   $-$0.29 \blocks{0}      & 1/2 \ce{Ag2Te} + 1/2   \ce{AgCl} \\
\ce{Ag4Cl2Te}    & 1.95     & 1.90  \blocks{0}      & \blocks{0}   $-$0.19 \blocks{0}      & 3/7 \ce{Ag2Te} + 4/7   \ce{AgCl} \\
\ce{Ag5Cl3Te}    & 2.24     & 1.99  \blocks{0}      & \blocks{0}   $-$0.10 \blocks{0}      & 1/3 \ce{Ag2Te} + 2/3   \ce{AgCl} \\
\ce{Ag4Cl2S}     & 2.26     & 1.53  \blocks{0}      & \blocks{0}  \ \ \  0.10  \blocks{2}      & 3/7 \ce{Ag2S} + 4/7   \ce{AgCl}  \\
\ce{Ag4Br2S}     & 2.04     & 1.45  \blocks{0}      & \blocks{0}  \ \ \  0.01  \blocks{1}      & 3/7 \ce{Ag2S} + 4/7   \ce{AgBr}  \\
\ce{Ag5Br3S}     & 2.23     & 1.61  \blocks{0}      & \blocks{0}  \ \ \  0.03  \blocks{1}      & 1/3 \ce{Ag2S} + 2/3   \ce{AgBr}  \\
\ce{Cl2Cu8S3}    & 1.88     & 1.14  \blocks{2}      & \blocks{0}   $-$0.17 \blocks{0}      & 1/3 \ce{CuCl} + 2/3   \ce{Cu2S}  \\
\ce{Cu10Cl4S3}   & 2.24     & 1.27  \blocks{0}      & \blocks{0}   $-$0.10 \blocks{0}      & 1/2 \ce{CuCl} + 1/2   \ce{Cu2S}  \\
\ce{PbCuSeCl}    & 1.84     & 1.23  \blocks{0}      & \blocks{0}   $-$0.25 \blocks{0}      & 1/2 \ce{CuCl} + 1/2   \ce{PbSe}  \\
\ce{Ti2O4Pb3Se3} & 1.64     & 1.26  \blocks{0}      & \blocks{0}   $-$0.72 \blocks{0}      & 1/2 \ce{PbSe} + 1/2   \ce{TiO2}  \\ \hline
\ce{Parents} & & & & \\ \hline
\ce{Ag2Te}       & 0.17     & 1.38  \blocks{0}      & \blocks{0}   $-$0.74 \blocks{0}      &                                  \\
\ce{AgCl}        & 3.28     & 2.30  \blocks{0}      & \blocks{0}  \ \ \  0.22  \blocks{0}      &                                  \\
\ce{Ag2S}        & 0.90     & 0.50  \blocks{0}      & \blocks{0}   $-$0.07 \blocks{0}      &                                  \\
\ce{AgBr}        & 2.89     & 2.16  \blocks{0}      & \blocks{0}  \ \ \  0.07  \blocks{0}      &                                  \\
\ce{Cu2S}        & 1.20     & 0.89  \blocks{0}      & \blocks{0}   $-$0.30 \blocks{0}      &                                  \\ \hline \hline
\end{tabularx}
\end{table*}


Figure~\ref{fig:InterpolationDesirablePropertySpace} provides insight into how certain interpolation patterns emerged as dominant. These patterns can be understood in relation to the initial distribution of compounds in property space.
Few initial compounds had acceptable $E_g$ or $\phi_{ox}$ and none had both simultaneously; however, acceptable $\phi_{red}$ was much more common. 
This combination advantageously positioned those with relatively high $\phi_{ox}$, especially the circled cluster containing the five highest $\phi_{ox}$ compounds, because many partners were located across the ideal region from them.
In fact, compounds from this cluster constituted one partner in nearly all interpolation pairings depicted in Figure~\ref{fig:InterpolationDesirablePropertySpace}, with the other partner being out-of-cluster and usually low $\phi_{red}$.

These pairings had the largest interpolation distance within the ideal region when the out-of-cluster partner was among the highest $\phi_{ox}$ of the low-$E_g$ compounds. Larger interpolation distance correlates with a greater number of possible hybrid compounds so this was the most dominant type of interpolation in our hybrid compound dataset. Thus we can roughly understand the interpolation opportunities available to our dataset by focusing on just a small subset of low-$E_g$ and high-$E_g$ compounds which are least oxidizable.

The four highest $\phi_{ox}$ compounds in the high-$E_g$ cluster were \ce{AgBr}, \ce{TiO2}, \ce{AgCl}, and \ce{CuCl}, ordered by the number of hybrids derived from them. 95\% of hybrid compounds had a parent in this group, including 42\% from \ce{AgBr} alone. The four highest $\phi_{ox}$ compounds with low-$E_g$ were the binary combinations of \ce{Pb} and \ce{Ag} with \ce{Se} and \ce{Te}. 40\% of hybrids had a parent in this group and 36\% that had one parent from each group. 
Table \ref{tab:SubMatrix_Compounds} provides some example hybrid compounds from the target region including three hybrids of different composition from pairs of \ce{AgBr} and \ce{TiO2}  with lower $E_g$ compounds. The variety of hybrids included represents how different parents produced hybrids in different regions, e.g. \ce{TiO2} -- \ce{PbO} hybrids tended to have low $E_g$.

\subsection{Candidate Structures}
We used the procedure for USPEX and VASP laid out in Section~\ref{sect:USPEXSettings} to search for the crystal structures of  hybrids in our target region. USPEX was able to converge structures for about 50 hybrid compounds. The elemental composition of these structures mostly coincides with the composition in the hybrid compounds highlighted in the previous Section. For example, \ce{Ag} has the greatest occurrence by far, due to its presence in both the low and high $E_g$ groups. However, \ce{Br} has a surprisingly much lower occurrence and \ce{Cd} has a relatively higher occurrence. Figure~\ref{fig:USPEX15structures} and Table~\ref{tab:USPEXStructs} show example results of USPEX converged structures. Figure~\ref{fig:USPEX15structures} also connects shifts in composition to changes in structure and property space.

USPEX allowed us to find hybrid structures that had different space groups from their parent structures. For example, PbCuSeCl has a space group of 156, which is different than any known PbSe or CuCl structure according to AFLOW. \citep{AFLOW} The approach of \citet{SubstitutionMatrix} and most other materials searches using the substitution matrix technique would not have allowed the discovery of this structure because they enforced a restriction new candidate compounds had the same crystal structure as compounds from which they were derived.

\subsection{Final Water-Splitters}

\begin{figure}[h]
    \centering
\includegraphics[width=1\textwidth]{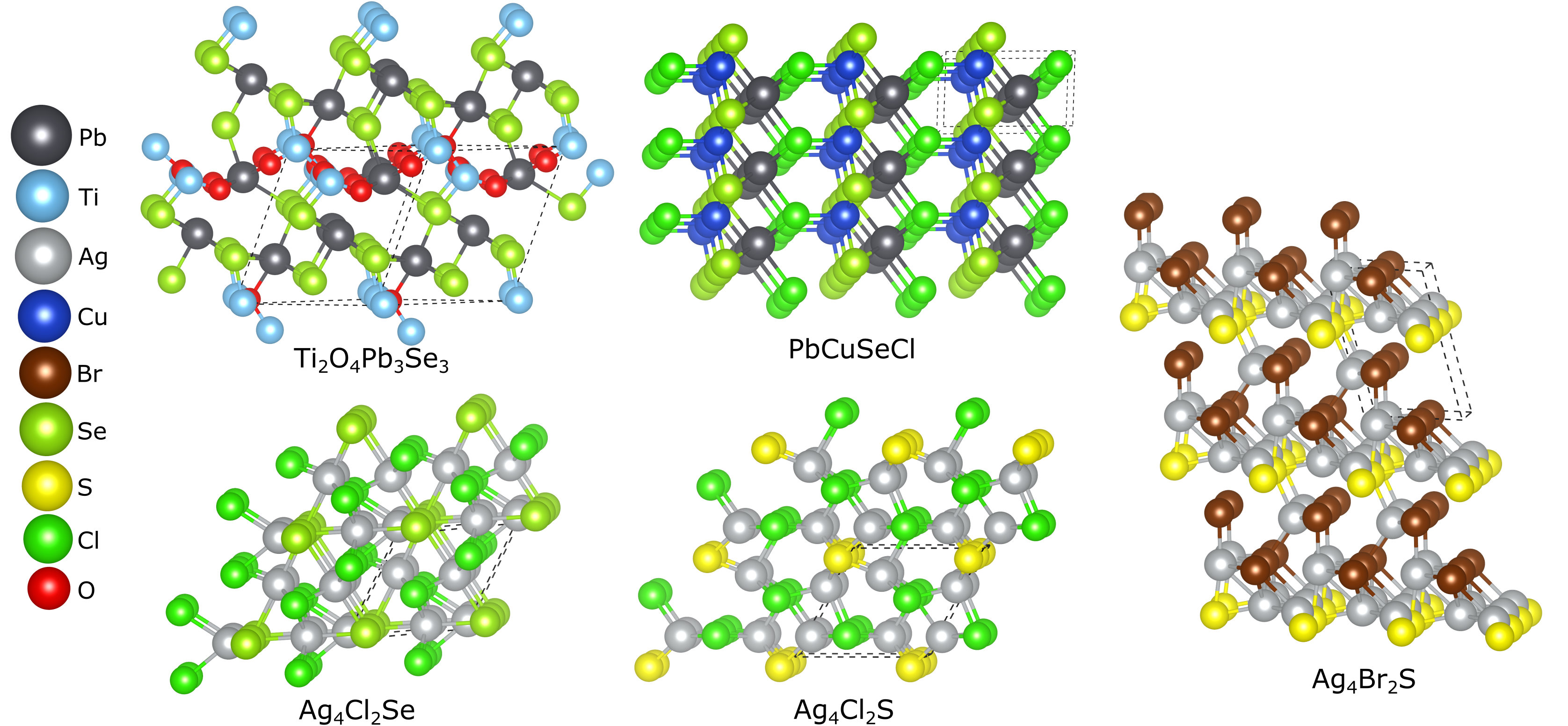}
    \caption{Crystal structures of final compounds with properties suitable for photocatalytic water-splitting. Atom sizes are consistent throughout figure for a given element, except atoms in the legend, which are 2 times as large in diameter. For each structure, dashed gray lines indicate the extent of a single conventional unit cell.} 
    \label{fig:FinalStructuresSALSA}
\end{figure}

Finally, we used the hybrid DFT code, CRYSTAL17, to conduct higher fidelity geometry optimization on our candidate structures. Many of the USPEX-converged structures had to be eliminated because the band gap derived from CRYSTAL17 moved them out of the target region of property space and others had to be eliminated because they would not converge. Structures containing \ce{Ag} had larger discrepancies between their interpolated $E_g$ and their CRYSTAL17-calculated $E_g$. On average the difference was 0.92~eV compared to 0.51~eV for non-\ce{Ag} compounds. The largest difference was for \ce{Ag4Cl2Se} which decreased by $1.17$ eV, placing it slightly out of the ideal region. The surviving structures mostly had low symmetry. 
Figure~\ref{fig:FinalStructuresSALSA} and Table~\ref{tab:FinalStructuresSALSA} provide examples of final structures that remained in the target region of property space after CRYSTAL17 band gap calculation. We confirmed that no final structure existed in the AFLOW database. \citep{AFLOW}

\clearpage
\newpage

\onecolumngrid

\begin{table}[h]
\centering
\caption{Final compounds with band gaps and redox potentials suitable for photocatalytic water-splitting. One \textmidscript{\ding{121}}-symbol appears next to a value for each 0.05~eV/V it lies outside of the ideal range (rounded down).} \label{tab:FinalStructuresSALSA}
\setlength{\tabcolsep}{11pt}
\begin{tabularx}{1\textwidth}{lccccc}
\hline \hline
Compound & Band Gap (eV) & Oxidation Potential (V) & Reduction Potential (V)  & Space Group & Price (USD/kg)\\ \hline
\ce{Ti2O4Pb3Se3 } & 2.333 \blocks{0}  & 1.257 \blocks{0}  &   \blocks{0} $-$0.717 \blocks{0} & 1 & 8\\
\ce{PbCuSeCl    } &1.512 \blocks{0} & 1.225 \blocks{1}   &  \blocks{0} $-$0.246 \blocks{0} & 156 & 7\\
\ce{Ag4Br2S    }  &2.741 \blocks{0}  & 1.451 \blocks{0}  &  \blocks{0}  \ \ \  0.014 \blocks{1} & 1 & 307\ \ \\
\ce{Ag4Cl2Se}     &1.058 \blocks{4}  & 1.907 \blocks{0}  &  \blocks{0} $-$0.007 \blocks{0} & 1 & 299\ \ \\
\ce{Ag4Cl2S   }   &1.060 \blocks{4} & 1.527 \blocks{0}   & \blocks{0}  \ \ \  0.099 \blocks{2} & 1 & 301\ \ \\ \hline \hline
\end{tabularx}
\end{table}

\twocolumngrid

\section{Discussion} \label{sect:Discussion}

\begin{figure}
    \centering
\includegraphics[width=1\columnwidth]{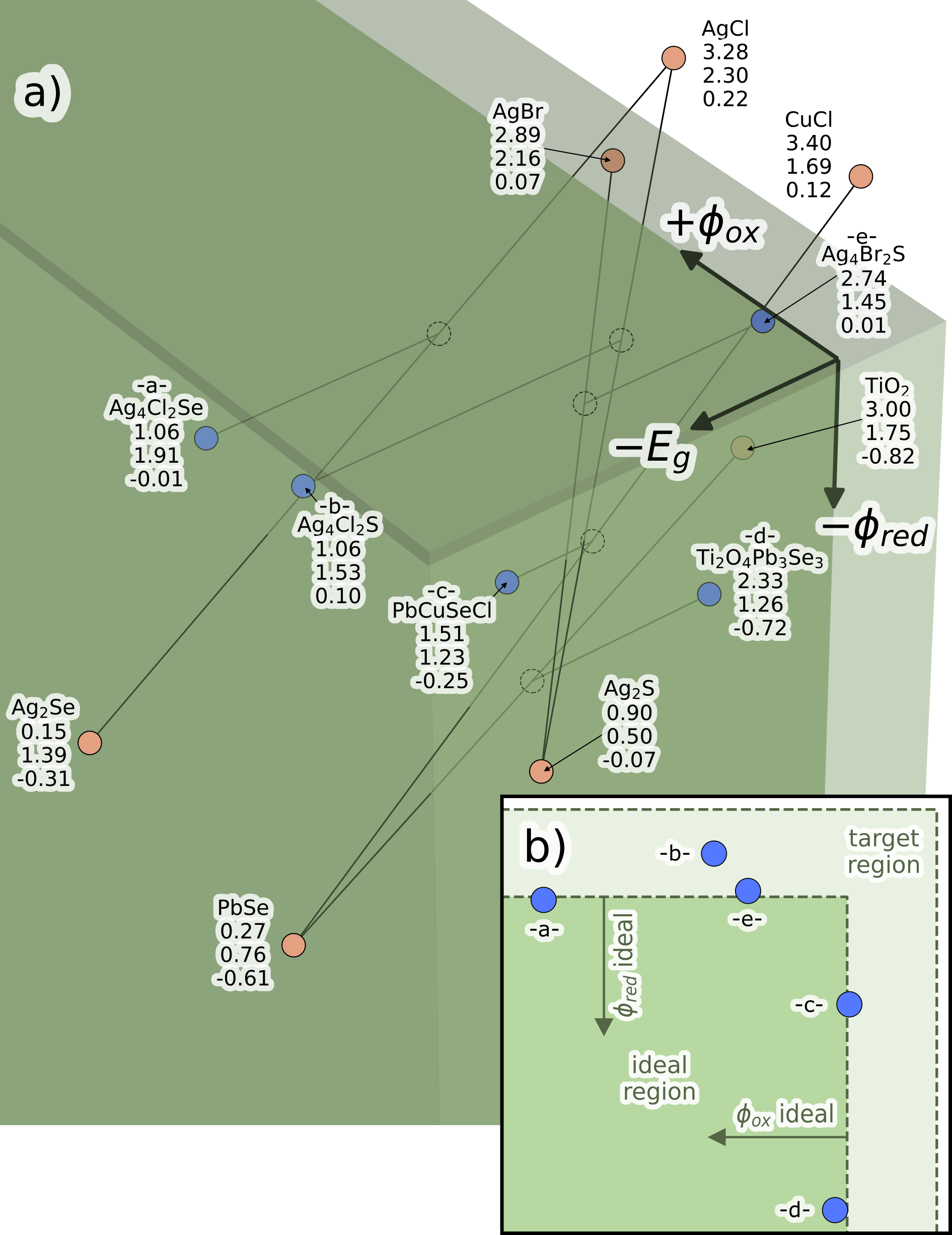}
    \caption{a) Property space diagram with undiscovered semiconductors with desirable band gaps (eV), oxidation potentials (V) and reduction potentials (V) produced by SALSA depicted in blue. Interpolated predictions depicted as unfilled circles. Original compounds are depicted as peach-colored points. b) A 2D $\phi_{ox}$-$\phi_{red}$ projection that demonstrates which boundaries final Pb and Ag compounds lie near. Final compounds are lettered in correspondance with a) to conserve space. This projection is 1.0~V$\times$1.0~V in extent.} 
    \label{fig:FinalInterpolationsPropertySpace}
\end{figure}

Figure~\ref{fig:FinalInterpolationsPropertySpace} (a) presents the interpolations into property space from our initial compounds which yielded our final structures, as well as shifts from interpolated predictions. Trends within this subset of interpolations suggest certain paths are favored for producing a photocatalytic water-splitter.

Our final materials can be divided into two groups. One group is made up of materials containing Silver, halides and group 16 compounds. Among the few compounds in our initial dataset which had good oxidation potentials, most contained Silver, so this group emerged from the interpolation between a pair of materials which had good oxidation potentials, but which had band gaps that were too low and too high respectively. Consequently, these materials are robust to hole oxidation -- all interpolated oxidation potentials are at least 0.2~V greater than the ideal minimum. However, their interpolated reduction potentials lie close to the threshold for rejection -- none are more than 0.01~V under the ideal maximum. Additionally, these structures have low symmetry and are expensive due to their Silver content.

The other group contains Lead instead of Silver. Redox suitability of these Lead compounds is inverted relative to the Silver group. That is, these compounds are robust to electron reduction due to Lead's especially negative reduction potential -- all have reduction potentials more than 0.2~V under the ideal maximum. However, none have an interpolated oxidation potential that is 0.03~V greater than the ideal minimum. The Lead structures are also higher in symmetry and relatively cheap. Figure~\ref{fig:FinalInterpolationsPropertySpace} (b) highlights how compounds from different groups lie near different planes of the desired property space, demonstrating the strengths and weaknesses of these groups.  

The Lead group is about 50 times cheaper so it may offer more scalability. \citep{ElementPrices1,ElementPrices2,ElementPrices3,ElementPrices4,ElementPrices5} However, the Silver group follows more closely to regular compositional formula. This means it may be easier to find more compounds in this group with different interpolation ratios if the ones we have discovered do not prove to be as effective as they appear to be. Both paths should be investigated experimentally.

Furthermore, we envision the materials design approach we used to be generalizable. In a different scheme, we would see a picture similar to this, but with different starting compounds and different boundaries than our target property space.

\clearpage 

\section{Methods} \label{sect:Methods}
\subsection{Initial Dataset} \label{sect:Inituak}

\begin{figure}
    \centering
\includegraphics[width=.99\columnwidth]{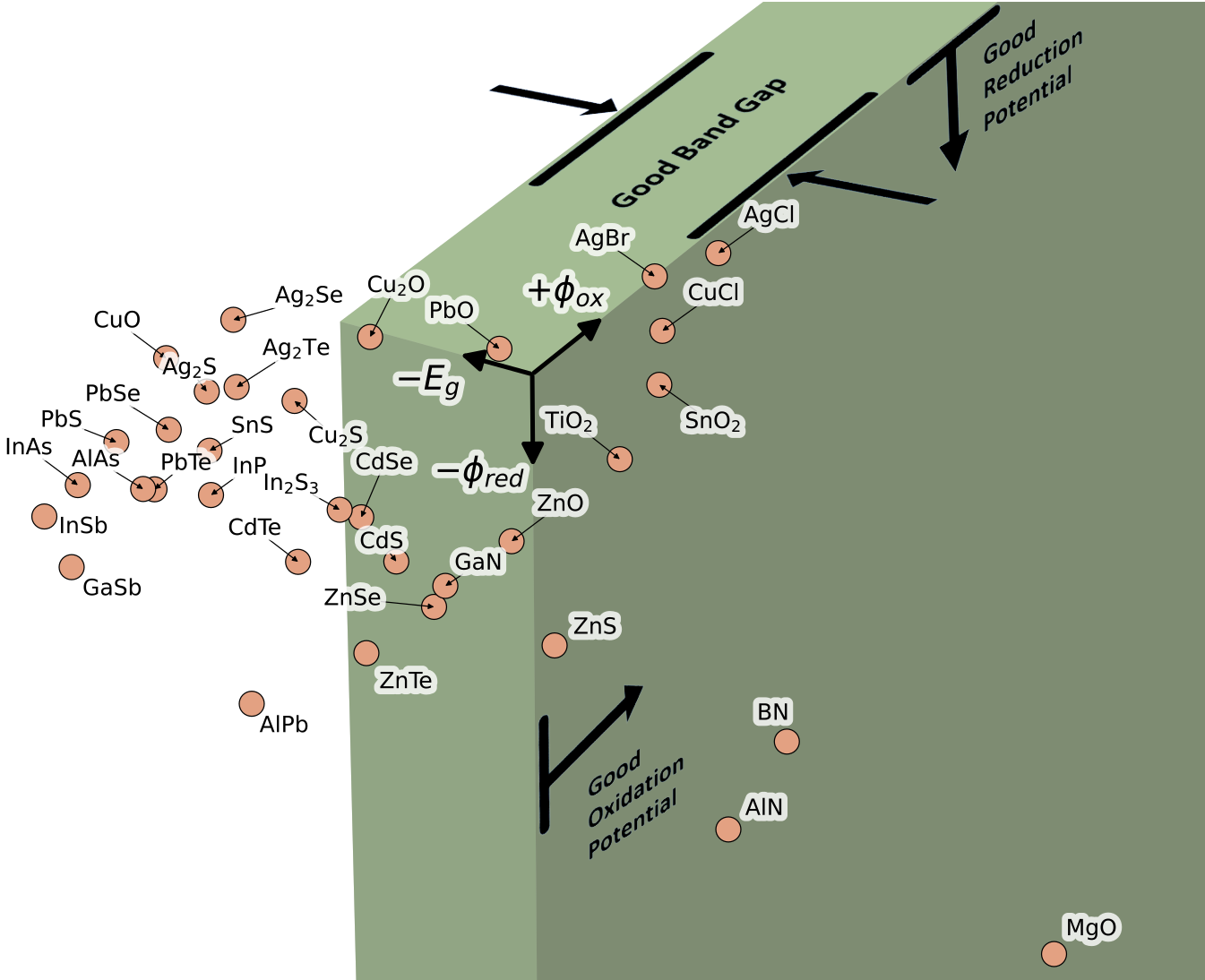}
    \caption{An overlook on the initial dataset of compounds used in our application of SALSA to artificial photosynthesis. Here they are depicted in a band gap -- oxidation potential -- reduction potential property space. The region of property space suitable for photocatalytic water-splitting is indicated. All 34 initial compounds are depicted.}
    \label{fig:InitialCompoundPropertySpace}
\end{figure}

We first collected a dataset of experimentally determined semiconductor band gaps. We then applied the method described in Stafford et al. 2023c in prep to calculate reduction and oxidation potentials. This formed an initial dataset containing 34 compounds. We sought compounds with band gaps between 1.23--2.80~eV to enable efficient photoabsorption. We also sought reduction potentials below 0.00 V and oxidation potentials above 1.23~V, with respect to the NHE, for materials which are stable in an aqueous environment. None of our original materials had suitable values for all three properties. Figure~\ref{fig:InitialCompoundPropertySpace} presents an overview of the collection of initial compounds and the region of property space described above. Table \ref{tab:InitialCompounds} lists each initial compound, its band gap and its redox potentials.

Figure~\ref{fig:PropertySpaceFromDifferentSides} in the Supplementary Material presents a closer look at the swarm of compounds that hover just outside of the target space. Few compounds are stable to photo-catalyzed decomposition. This is mainly because most have oxidation potentials that are too low, leaving them prone to hole oxidation (Figure~\ref{fig:PropertySpaceFromDifferentSides} (d)).
However, some of the few with acceptable oxidation potentials have reduction potentials that are too high (Figure~\ref{fig:PropertySpaceFromDifferentSides} (b)). Additionally, compounds are roughly evenly divided into three groups which cannot absorb sunlight efficiently, cannot absorb it at all in the regions of higher solar intensity, and which have an acceptable band gap (Figure~\ref{fig:PropertySpaceFromDifferentSides} (e) and (c)). No matter which angle we look at this property space, we see there is great room for improvement.

\begin{table}
\centering
\caption{The compounds of our initial dataset with their known band gaps and redox potentials. \checkmark, \textdownarrow, and \textuparrow \ symbols indicate whether property values are suitable, too low or too high for photocatalytic water splitting, respectively.} \label{tab:InitialCompounds}
\setlength{\tabcolsep}{6pt}
\begin{tabularx}{0.48\textwidth}{lccc}
\hline \hline
Compound        & Band gap  (eV) & Oxidation (V) & Reduction (V) \\  \hline
Figure~\ref{fig:FinalStructuresSALSA} & & & \\ \hline
\ce{Ag2S}  & 0.90 \   $\midscript{\downarrow}$ \ \ \ & \ \ \ 0.50 \ $\midscript{\downarrow}$ \ \ \ & $-$0.07 \,$\midscript{\checkmark}$ \ \  \\
\ce{Ag2Se} & 0.15 \ $\midscript{\downarrow}$ \ \ \   & \ \ \ 1.39 \,$\midscript{\checkmark}$ \ \   & $-$0.31 \,$\midscript{\checkmark}$ \ \  \\
\ce{Ag2Te} & 0.17 \ $\midscript{\downarrow}$ \ \ \   & \ \ \ 1.38 \,$\midscript{\checkmark}$ \ \   & $-$0.74 \,$\midscript{\checkmark}$ \ \  \\
\ce{AgBr}  & 2.89 \ $\midscript{\uparrow}$ \ \ \     & \ \ \ 2.16 \,$\midscript{\checkmark}$ \ \   & \ \ \ 0.07 \ $\midscript{\uparrow}$ \ \ \ \\
\ce{AgCl}  & 3.28 \ $\midscript{\uparrow}$ \ \ \     & \ \ \ 2.30 \,$\midscript{\checkmark}$ \ \   & \ \ \ 0.22 \ $\midscript{\uparrow}$ \ \ \ \\
\ce{CuCl}  & 3.40 \ $\midscript{\uparrow}$ \ \ \     & \ \ \ 1.69 \,$\midscript{\checkmark}$ \ \   & \ \ \ 0.12 \ $\midscript{\uparrow}$ \ \ \ \\
\ce{PbSe}  & 0.27 \ $\midscript{\downarrow}$ \ \ \   & \ \ \ 0.76 \ $\midscript{\downarrow}$ \ \ \ & $-$0.61 \,$\midscript{\checkmark}$ \ \  \\ 
\ce{TiO2}  & 3.00 \ $\midscript{\uparrow}$ \ \ \     & \ \ \ 1.75 \,$\midscript{\checkmark}$ \ \   & $-$0.83 \,$\midscript{\checkmark}$ \ \  \\ \hline
Other & & & \\ \hline
\ce{AlAs}  & 2.20 \,$\midscript{\checkmark}$ \ \     & $-$1.11 \ $\midscript{\downarrow}$ \ \ \  & \ \ \ 0.64 \ $\midscript{\uparrow}$ \ \ \ \\
\ce{AlN}   & 6.00 \ $\midscript{\uparrow}$ \ \ \     & $-$0.53 \ $\midscript{\downarrow}$ \ \ \  & $-$0.90 \,$\midscript{\checkmark}$ \ \  \\
\ce{AlPb}  & 2.80 \ $\midscript{\uparrow}$ \ \ \     & $-$0.94 \ $\midscript{\downarrow}$ \ \ \  & $-$0.62 \,$\midscript{\checkmark}$ \ \  \\
\ce{BN}    & 6.00 \ $\midscript{\uparrow}$ \ \ \     & $-$0.06 \ $\midscript{\downarrow}$ \ \ \  & $-$0.70 \,$\midscript{\checkmark}$ \ \  \\
\ce{CdS}   & 2.58 \,$\midscript{\checkmark}$ \ \     & \ \ \ 0.35 \ $\midscript{\downarrow}$ \ \ \ & $-$0.67 \,$\midscript{\checkmark}$ \ \  \\
\ce{CdSe}  & 1.85 \,$\midscript{\checkmark}$ \ \     & \ \ \ 0.78 \ $\midscript{\downarrow}$ \ \ \ & $-$0.83 \,$\midscript{\checkmark}$ \ \  \\
\ce{CdTe}  & 1.61 \,$\midscript{\checkmark}$ \ \     & \ \ \ 0.51 \ $\midscript{\downarrow}$ \ \ \ & $-$0.99 \,$\midscript{\checkmark}$ \ \  \\
\ce{Cu2O}  & 2.10 \,$\midscript{\checkmark}$ \ \     & \ \ \ 0.64 \ $\midscript{\downarrow}$ \ \ \ & \ \ \ 0.44 \ $\midscript{\uparrow}$ \ \ \ \\
\ce{Cu2S}  & 1.20 \ $\midscript{\downarrow}$ \ \ \   & \ \ \ 0.89 \ $\midscript{\downarrow}$ \ \ \ & $-$0.30 \,$\midscript{\checkmark}$ \ \  \\
\ce{CuO}   & 1.20 \ $\midscript{\downarrow}$ \ \ \   & $-$0.05 \ $\midscript{\downarrow}$ \ \ \  & \ \ \ 0.54 \ $\midscript{\uparrow}$ \ \ \ \\
\ce{GaN}   & 3.40 \ $\midscript{\uparrow}$ \ \ \     & $-$0.06 \ $\midscript{\downarrow}$ \ \ \  & $-$0.36 \,$\midscript{\checkmark}$ \ \  \\
\ce{GaSb}  & 0.73 \ $\midscript{\downarrow}$ \ \ \   & $-$0.38 \ $\midscript{\downarrow}$ \ \ \  & $-$0.65 \,$\midscript{\checkmark}$ \ \  \\
\ce{In2S3} & 1.98 \,$\midscript{\checkmark}$ \ \     & \ \ \ 0.49 \ $\midscript{\downarrow}$ \ \ \ & $-$0.57 \,$\midscript{\checkmark}$ \ \  \\
\ce{InAs}  & 0.41 \ $\midscript{\downarrow}$ \ \ \   & $-$0.03 \ $\midscript{\downarrow}$ \ \ \  & $-$0.42 \,$\midscript{\checkmark}$ \ \  \\
\ce{InP}   & 1.42 \,$\midscript{\checkmark}$ \ \     & \ \ \ 0.05 \ $\midscript{\downarrow}$ \ \ \ & $-$0.31 \,$\midscript{\checkmark}$ \ \  \\
\ce{InSb}  & 0.23 \ $\midscript{\downarrow}$ \ \ \   & $-$0.13 \ $\midscript{\downarrow}$ \ \ \  & $-$0.60 \,$\midscript{\checkmark}$ \ \  \\
\ce{MgO}   & 7.80 \ $\midscript{\uparrow}$ \ \ \     & \ \ \ 0.18 \ $\midscript{\downarrow}$ \ \ \ & $-$1.73 \,$\midscript{\checkmark}$ \ \  \\
\ce{PbO}   & 2.70 \,$\midscript{\checkmark}$ \ \     & \ \ \ 1.07 \ $\midscript{\downarrow}$ \ \ \ & \ \ \ 0.24 \ $\midscript{\uparrow}$ \ \ \ \\
\ce{PbS}   & 0.37 \ $\midscript{\downarrow}$ \ \ \   & \ \ \ 0.29 \ $\midscript{\downarrow}$ \ \ \ & $-$0.37 \,$\midscript{\checkmark}$ \ \  \\
\ce{PbTe}  & 0.32 \ $\midscript{\downarrow}$ \ \ \   & \ \ \ 0.60 \ $\midscript{\downarrow}$ \ \ \ & $-$0.88 \,$\midscript{\checkmark}$ \ \  \\
\ce{SnO2}  & 3.50 \ $\midscript{\uparrow}$ \ \ \     & \ \ \ 1.56 \,$\midscript{\checkmark}$ \ \   & $-$0.12 \,$\midscript{\checkmark}$ \ \  \\
\ce{SnS}   & 1.00 \ $\midscript{\downarrow}$ \ \ \   & \ \ \ 0.42 \ $\midscript{\downarrow}$ \ \ \ & $-$0.37 \,$\midscript{\checkmark}$ \ \  \\
\ce{ZnO}   & 3.37 \ $\midscript{\uparrow}$ \ \ \     & \ \ \ 0.48 \ $\midscript{\downarrow}$ \ \ \ & $-$0.45 \,$\midscript{\checkmark}$ \ \  \\
\ce{ZnS}   & 3.84 \ $\midscript{\uparrow}$ \ \ \     & \ \ \ 0.35 \ $\midscript{\downarrow}$ \ \ \ & $-$0.90 \,$\midscript{\checkmark}$ \ \  \\
\ce{ZnSe}  & 2.83 \ $\midscript{\uparrow}$ \ \ \     & \ \ \ 0.40 \ $\midscript{\downarrow}$ \ \ \ & $-$0.93 \,$\midscript{\checkmark}$ \ \  \\
\ce{ZnTe}  & 2.39 \,$\midscript{\checkmark}$ \ \     & \ \ \ 0.29 \ $\midscript{\downarrow}$ \ \ \ & $-$1.25 \,$\midscript{\checkmark}$ \ \  \\ \hline \hline
\end{tabularx}
\end{table}

\subsection{Parameters Used for Candidate Generation} \label{sect:CandidateParameters}
\paragraph{Substitution Threshold}
We used a substitution threshold of 0, that is, we did not consider substitutions associated with negative values in our substitution likelihood matrix. This parameter can be adjusted as governed by the computational resources available to a search. A lower threshold enables a more thorough exploration of composition space, but is more computationally expensive and less efficient at finding suitable materials.
 
\paragraph{Substitution Implementation}
We allowed substitution to constitute a complete or partial replacement of the original ion. For example, $\ce{Br^-}\leftrightarrow\ce{I^-}$ is a matrix-allowed substitution and \ce{AgBr} is in our initial dataset so compounds of the form \ce{Ag_n Br_{n-m} I_m} with  $n, m \in \mathbb{Z}$, are in our candidate dataset. We limited substitutions to be first or second order, i.e at most two substitutions could be used to generate an individual candidate. In Section \ref{sect:SALSAapplied}, first and second-order substitutions correspond to ternary and quaternary compounds, respectively. Theoretically, a second-order substitution could consist of exchanging a single, original ionic component for two new ions. However, second-order substitutions that formed hybrid compounds consisted of a single substitution of each of the original components as this is the only way a second-order substitution could correspond to interpolation between two binary compounds.
Building on the previous example, the substitution $\ce{Ag^+}\leftrightarrow\ce{Cu^+}$ could be used in second-order substitutions to produce quaternary compounds of the form \ce{Ag_{n-p} Cu_p Br_{n-m} I_m} with $n, p, m \in \mathbb{Z}$. For the purpose of enumerating a complete dataset, the new components of candidate compounds were limited to half or less the final composition. So all second-order substitutions were partial and you could not generate \ce{Cu I} ($n=p=m=1$) from just \ce{AgBr}. However, this limitation does not affect hybrid compounds.

\paragraph{Unit Cell Size Limit}
In practice, the enumeration of candidate compounds requires some constraint on values of $n, m,p$. Results presented in Section \ref{sect:SALSAapplied} implemented this constraint by imposing a maximum of 20 atoms in a unit cell. This is equivalent to the constraints 
$1 \le n,p,m \le 10$ in the previous example.

\subsection{Property Space Selection Criteria} \label{sect:SelectCrit}
With our interpolation scheme we filtered compounds that did not meet the following criteria: 1.03 $<$  band gap (eV) $<$ 3.00, oxidation potential (V) $>$ 1.03, and reduction potential (V) $<$ 0.2.
This includes an extra window of 0.20~eV for the band gaps and 0.20~V for the potentials to allow for materials that might ultimately arrive in the desired region of property space by deviating slightly from their linear interpolation. To illustrate this process, consider PbSe and CuCl. PbSe's band gap is too small, at 0.27~eV, and its oxidation potential is too low, at 0.76~eV, while CuCl has too high a band gap at 3.40~eV. However, the 50:50 interpolation between these two, PbCuSeCl, has band gap, oxidation and reduction potentials of 1.84~eV, 1.23~V and $-$0.25~V, respectively, which places it just inside the threshold of our target region.

\subsection{Interpolation} \label{sect:Interpolation}
We construct hybrid compositions which are integer ratios of two parent compositions. We then estimate the properties of the corresponding hybrid compounds to be linear interpolations of the parent compounds on a per-atom basis. In other words, we weight the initial property values by the number of atoms contributed to the hybrid. 
Furthermore, we don't restrict our interpolations to be single-substitution. For example, both $\ce{Pb^2+}\leftrightarrow~\ce{Cu+}$ and $\ce{Se^2-}\leftrightarrow~\ce{Cl-}$ are matrix-allowed substitutions so if we start with initial compounds PbSe and CuCl, we generate interpolated compositions such as PbCuSeCl.

To better understand the per-atom weighting procedure, consider an illustrative example in which we have initial compounds \ce{Ag2S} and \ce{AgBr} that have a property with values $0$ and $P$. $\ce{S^2-}\leftrightarrow\ce{Br^-}$ is a matrix-allowed substitution, so we consider composition ratios of \ce{Ag2S} and \ce{AgBr} such as 2:1, 1:1, and 1:2, which correspond to \ce{Ag5BrS2}, \ce{Ag3BrS}, and \ce{Ag4Br2S}, respectively. According to our interpolation procedure, these new candidate compounds have estimated property values of $\frac{4}{7} P$, $\frac{2}{5} P$, and $\frac{1}{4} P$, respectively. Note that \ce{Ag3BrS} has a property value of $\frac{2}{5} P$ rather than $\frac{1}{2} P$, despite being a 1:1 ratio of initial compositions. To understand this potentially nonintuitive result, recognize that 2 of the 5 atoms in \ce{Ag3BrS} were contributed by \ce{AgBr} so its interpolation weight is $\frac{2}{5}$ and accordingly, the interpolation weight of \ce{Ag2S} is $\frac{3}{5}$. Therefore, $P_{new} = \frac{2}{5} \times P + \frac{3}{5} \times 0 = \frac{2}{5}P$
 
\subsection{USPEX Settings} \label{sect:USPEXSettings}
We provide USPEX a composition and allow it to perform all stochastic modifications it has at its disposal. We do not constrain the structure by space group. For energy evaluation, we elect USPEX's option to interface with the DFT code, Vienna Ab initio Simulation Package (VASP).\citep{VASP1, VASP2, VASP3} All VASP calculations were performed in the plane-wave DFT framework at the Generalized Gradient Approximation (GGA) level of theory and used the Perdew, Burke, and Ernzerhof (PBE) functional. \citep{PBE} Projector-augmented wave (PAW) pseudopotentials were used to represent the core electrons and ion-electron interactions. \citep{VASPPOTCARs, VASPPOTCARs2}
We used a plane-wave cutoff of 500 eV, an energy convergence criterion of $10^{-4}$~eV, and force convergence of 0.02~eV/\AA. Dispersive interactions were accounted for using DFT-D3 corrections \cite{D3} with Becke-Jonson damping.\citep{BJ_damping} We also included spin polarization effects.

\subsection{CRYSTAL17 Settings} \label{sect:CRYSTALSettings}
We used the hybrid DFT code CRYSTAL17 to conduct higher fidelity geometry optimization on our candidate structures. CRYSTAL17 uses basis sets of atom-centered Gaussian-type functions. \citep{CRYSTAL17, CRYSTAL17_2}
We used the hybrid Heyd–Scuseria–Ernzerhof (HSE06) functional. \citep{HSE06_2003, HSE06_2006}
We also considered spin-polarization effects and used relativistic compact effective potentials and efficient, shared-exponent basis sets. \citep{Stevens1984BSs, Stevens1992BSs} The effective potentials were used for O, Cu, Se, Ag, Te and Pb. 
We included full geometry optimization of cell positions and lattice parameters. 
We sampled the reciprocal space for all the structures using a $\Gamma$-centered Monkhorst-Pack scheme with a resolution of 
$a_i n_{k_i} \ge 40\ \AA$ where $a_i$ and $n_{k_i}$ represent a lattice constant along the $i^{th}$ axis in real space and the number of k-points along the $i^{th}$ axis in reciprocal space, respectively. 
We optimized geometry with an SCF energy convergence criterion of $2.72 \times 10^{-6}$~eV, an RMS force criterion of $1.54 \times 10^{-2}$~eV/\AA, a max force criterion of $2.31 \times 10^{-2}$~eV/\AA, an RMS displacement of $6.35 \times 10^{-4}$~\AA, a max displacement criterion of $9.53 \times 10^{-4}$~\AA~and a between-geometry energy convergence criterion of $2.72 \times 10^{-6}$~\AA. 
For this application we also performed a single-point SCF optimization on the converged geometry to acquire a band gap, although this is not necessary for the SALSA workflow in general.  

\section{Conclusions} \label{sect:Conclusions}

We have introduced a general materials design process that can be used for many applications.
The process only requires a dataset of known compounds with known properties and the ability to calculate some of the properties from first-principles for a small set of structures.
We applied our new process to an unrealized artificial photosynthesis technology and were able to discover materials that are good candidates for photocatalytic water-splitting. This includes PbCuSeCl, a material with a novel structure, which we were able to discover because our process allows for an expansive search of structure space. It also includes \ce{Ti2O4Pb3Se3} which has band gap and interpolated redox potentials within the ideal range for photocatalytic water-splitting.

Furthermore, work is underway to improve several methods used in the SALSA process. We may expand and enhance further the substitution matrix. We are also working on a way to generalize the redox potential calculation method with larger datasets.

\begin{acknowledgments}
SMS is supported by the Mendoza Lab start-up funds. JLMC acknowledges start-up funds from Michigan State University. This work was supported in part by computational resources and services provided by the Institute for Cyber-Enabled Research at Michigan State University.

\textbf{Author Contributions}. AA and JLMC started the project in 2012-2013. JLMC conceived the idea and executed the first iterations of the search algorithms. AA and JLMC wrote the first draft. AA and JLMC implemented and developed the first iteration of the algorithms. SMS, MD, YL continued and finished the project. SMS implemented the next generation of the algorithm. 
Conceptualization: AA, JLMC. Methodology: AA, SMS, MD, YL, JLMC. Software: AA, SMS, MD, YL, JLMC. Validation: AA, SMS, MD, YL, JLMC. Formal Analysis: SMS, MD, JLMC. Investigation: AA, SMS, MD, JLMC. Resources: JLMC. Writing---original draft preparation: AA, JLMC. Writing---review and editing: SMS, AA, MD, YL, JLMC. Visualization: SMS, MD, JLMC. Supervision: JLMC. Project administration: JLMC.  Funding Acquisition: JLMC. All authors have read and agreed to the published version of the manuscript.

\end{acknowledgments}

\section*{Data Availability Statement}

The data that support the findings of this study are available from the corresponding author upon reasonable request.

\bibliography{myrefs.bib}

\newpage
\clearpage

\onecolumngrid

\begin{center}
	{\Huge  Supporting Information}
\end{center}
\vspace{3pt}

\hspace{-22pt} \rule{1.0\textwidth}{2.5pt}

\begin{center}
	
	\Large{\textbf{Transforming Materials Discovery for Artificial Photosynthesis: High-Throughput Screening of Earth-Abundant Semiconductors}} \\ [0.5cm]
	
	\Large
	Sean M. Stafford$^{1}$, Alexander Aduenko$^{2}$, Marcus Djokic$^{1}$, Yu-Hsiu Lin$^{1}$, Jose L. Mendoza-Cortes$^{1*}$ 
	\vspace{20pt}
	
	\normalsize
	
	$^{1}$ Department of Chemical Engineering and Material Science, Michigan State University, East Lansing, MI, 48824, USA.\\
	$^{2}$ Moscow Institute of Physics and Technology, Moscow, Russia.
	\vspace{3pt}
	
	\hspace{-22pt} \rule{1.0\textwidth}{2.5pt}
	
	\begin{center}
		\large
		Email: jmendoza@msu.edu
	\end{center}
	
\end{center}

\appendix*
\section{Supplementary Material}

\subsection{Hybrid Compound Enumeration and Max Unit Cell Size }
We found that the number of hybrid compounds for a given unit cell size, $n$, varied roughly linearly with $n$, at least for $4 \le n_{max} \le 50$.  About $70\ n$ hybrids of size $n$ could be generated, so the cumulative number of hybrids for $n_{max}$ was roughly $35\ n_{max}^2$. Constraining the hybrids to the target region and ideal region respectively, limited this to $6\ n$ and $2\ n + 3$ hybrids of size $n$ and therefore, $3\ n_{max}^2$ and $1\ n_{max}^2 + 3\ n_{max}$ cumulatively.

\subsection{Interpolation Distance Analysis}

The four highest $\phi_{ox}$ compounds with low-$E_g$ and high-$E_g$, respectively are \ce{Ag2Te}, \ce{Ag2Se}, \ce{PbSe}, and \ce{PbTe}; and \ce{AgBr}, \ce{TiO2}, \ce{AgCl}, and \ce{CuCl}, ordered by their total  interpolation distance.  These eight account for nearly 70\% of interpolation distance. 
\ce{SnO2}, \ce{Cu2S}, and \ce{PbS} also make significant, albeit lesser contributions to interpolation.

\subsection{Miscellaneous Additional Candidate Compound Data}

\begin{table}[h]
\centering
\caption{A selection of candidate compounds including titanium cuprates, tin cuprates and tin-lead oxides. These have interpolated band gaps and redox potentials which all lie within or close to their ideal ranges. One \textmidscript{\ding{121}}-symbol appears next to a value for each 0.05~eV/V it lies outside of the ideal range (rounded down).} \label{tab:SubMatrixCompounds2}
\setlength{\tabcolsep}{6pt}
\begin{tabularx}{0.48\textwidth}{lccc}
\hline \hline
Compound         & Band gap  (eV) & Oxidation (V) & Reduction (V) \\  \hline
\ce{Ti2Cu1O5}    & \blocks{0} 2.64 \blocks{0}       & \blocks{0} 1.39 \blocks{0}        & \blocks{0} $-$0.55   \blocks{0}      \\
\ce{Ti7Cu6O20}   & \blocks{0} 2.46 \blocks{0}       & \blocks{0} 1.21 \blocks{1}        & \blocks{0} $-$0.41   \blocks{0}      \\
\ce{Ti9Cu2O20}   & \blocks{0} 2.82 \blocks{1}       & \blocks{0} 1.57 \blocks{0}        & \blocks{0} $-$0.69   \blocks{0}   \\  
\ce{Ti3Cu4O10}   & \blocks{0} 2.28 \blocks{0}       & \blocks{0} 1.03 \blocks{4}        & \blocks{0} $-$0.28   \blocks{0}      \\ \hline
\ce{Sn7Cu6O20}   & \blocks{0} 2.81 \blocks{1}       & \blocks{0} 1.08 \blocks{3}        & \blocks{0}  \ \ \   0.08    \blocks{2}  \\
\ce{Sn3Cu8O10}   & \blocks{0} 2.94 \blocks{3}       & \blocks{0} 1.19 \blocks{1}        & \blocks{0} \ \ \   0.10   \blocks{2}    \\
\ce{Sn1Cu4O4}    & \blocks{0} 2.80 \blocks{0}       & \blocks{0} 1.10 \blocks{3}        & \blocks{0} \ \ \    0.16   \blocks{4}    \\
\ce{Sn3Pb14O20}  & \blocks{0} 2.94 \blocks{3}       & \blocks{0} 1.22 \blocks{1}        & \blocks{0} \ \ \   0.13  \blocks{3}        \\
\ce{Sn1Pb8O10}   & \blocks{0} 2.86 \blocks{2}       & \blocks{0} 1.17 \blocks{2}        & \blocks{0} \ \ \  0.17    \blocks{4}     \\ \hline \hline
\end{tabularx}
\end{table}

\subsection{Price Calculation}
We estimated a price (USD/kg) for final structures in Table \ref{tab:FinalStructuresSALSA} by performing a weighted sum of component elements prices. The component elements prices were weighted by fractional composition of the final structures. These elemental prices were found in technical reports. \citep{ElementPrices1,ElementPrices2,ElementPrices3,ElementPrices4,ElementPrices5}

\onecolumngrid

\subsection{Initial Compound Distribution}

\begin{figure}[h]
    \centering
\includegraphics[width=.91\columnwidth]{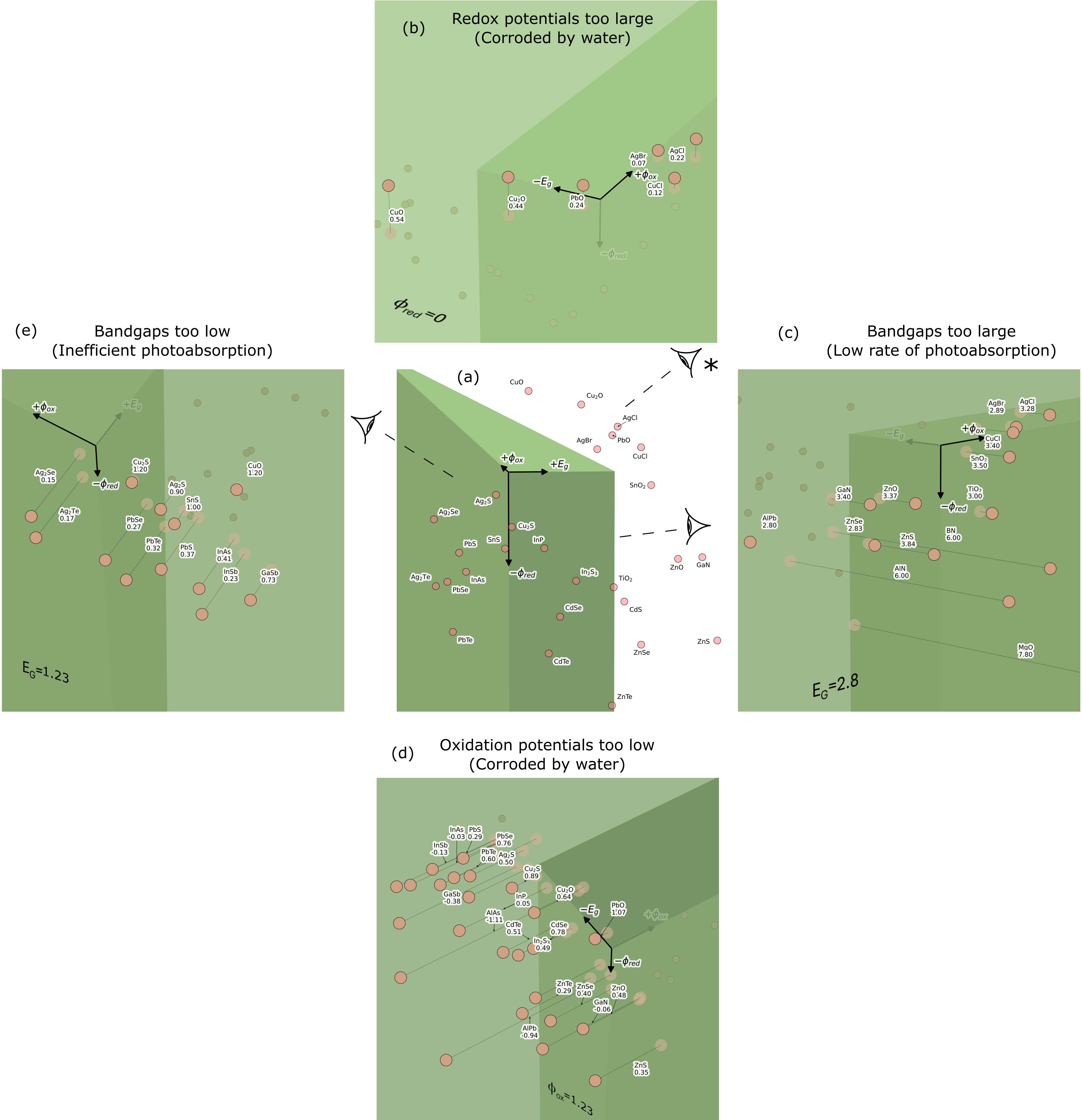}
    \caption{ A closer look at the obstacles to achieving photo-catalytic water-splitting as regions of property space. Part (a) provides an overview. 
    Parts (b)-(e) partition the initial band gap-redox space into regions of problematic qualities. Points representing compounds that fall into this segment are emphasized. They are projected onto the segmenting plane and the offending values are provided where space allows. Perspective is indicated by the eye symbol by the diagram. * (d) uses approximately the same perspective as (a). Parts (b) and (d) look at redox potentials that are too high and too low, respectively. Parts (c) and (e) look at band gaps that are too large and too small, respectively.}
    \label{fig:PropertySpaceFromDifferentSides}
\end{figure}


\clearpage
\newpage

\subsection{CIFs for Optimized Final Structures}

\begin{verbatim}
data_Ag4Cl2Se_ZmODi_HSE.out

_cell_length_a                         7.369297
_cell_length_b                         4.687832
_cell_length_c                         5.842447
_cell_angle_alpha                      85.904266
_cell_angle_beta                       109.294977
_cell_angle_gamma                      50.551477
_symmetry_space_group_name_H-M         'P 1'
_symmetry_Int_Tables_number            1

loop_
_symmetry_equiv_pos_as_xyz
   'x, y, z'

loop_
   _atom_site_label
   _atom_site_type_symbol
   _atom_site_fract_x
   _atom_site_fract_y
   _atom_site_fract_z
Ag001  Ag  -0.006202  -0.245962   0.106597
Ag002  Ag   0.289404   0.032758  -0.416624
Ag003  Ag   0.332202  -0.084534   0.105550
Ag004  Ag  -0.188334   0.360196  -0.371861
Cl005  Cl   0.233241  -0.437713  -0.225149
Cl006  Cl  -0.435779   0.134411   0.436497
Se007  Se  -0.180832   0.428445   0.105490

# end of cif
\end{verbatim}

\clearpage
\newpage
\begin{verbatim}
data_Ag4Br2S_Vjwqo_HSE.out

_cell_length_a                         4.735259
_cell_length_b                         4.881917
_cell_length_c                         7.874516
_cell_angle_alpha                      108.513975
_cell_angle_beta                       72.670342
_cell_angle_gamma                      90.197111
_symmetry_space_group_name_H-M         'P 1'
_symmetry_Int_Tables_number            1

loop_
_symmetry_equiv_pos_as_xyz
   'x, y, z'

loop_
   _atom_site_label
   _atom_site_type_symbol
   _atom_site_fract_x
   _atom_site_fract_y
   _atom_site_fract_z
Ag001  Ag   0.447157  -0.047077  -0.304960
Ag002  Ag  -0.343787   0.481097   0.286775
Ag003  Ag   0.150931   0.098349   0.296495
Ag004  Ag  -0.213091  -0.196346   0.015188
Br005  Br  -0.032808  -0.348144  -0.347098
Br006  Br   0.288539   0.401955   0.013198
 S007   S  -0.385439   0.024666   0.368502

# end of cif
\end{verbatim}

\clearpage
\newpage
\begin{verbatim}
data_Ag4Cl2S_VOn3j_HSE.out

_cell_length_a                         7.109721
_cell_length_b                         3.834986
_cell_length_c                         7.849064
_cell_angle_alpha                      55.452895
_cell_angle_beta                       126.564972
_cell_angle_gamma                      110.032698
_symmetry_space_group_name_H-M         'P 1'
_symmetry_Int_Tables_number            1

loop_
_symmetry_equiv_pos_as_xyz
   'x, y, z'

loop_
   _atom_site_label
   _atom_site_type_symbol
   _atom_site_fract_x
   _atom_site_fract_y
   _atom_site_fract_z
Ag001  Ag   0.026908  -0.462006  -0.341242
Ag002  Ag  -0.066125  -0.378441   0.205729
Ag003  Ag   0.350499   0.185052   0.191618
Ag004  Ag  -0.348568   0.469547  -0.237964
Cl005  Cl   0.340014   0.043477  -0.436151
Cl006  Cl  -0.340758  -0.107307   0.221834
 S007   S   0.044630  -0.230422  -0.074723

# end of cif
\end{verbatim}

\clearpage
\newpage
\begin{verbatim}
data_PbCuSeCl_3ioP2_HSE.out

_cell_length_a                         4.084009
_cell_length_b                         4.084009
_cell_length_c                         6.110847
_cell_angle_alpha                      90.000000
_cell_angle_beta                       90.000000
_cell_angle_gamma                      120.000000
_symmetry_space_group_name_H-M         'P 3 m 1'
_symmetry_Int_Tables_number            156

loop_
_symmetry_equiv_pos_as_xyz
   'x, y, z'

loop_
   _atom_site_label
   _atom_site_type_symbol
   _atom_site_fract_x
   _atom_site_fract_y
   _atom_site_fract_z
Pb001  Pb  -0.333333   0.333333   0.464329
Cu002  Cu   0.333333  -0.333333  -0.163608
Se003  Se   0.000000   0.000000  -0.253155
Cl004  Cl   0.333333  -0.333333   0.216834

# end of cif
\end{verbatim}

\clearpage
\newpage
\begin{verbatim}
data_Ti2O4Pb3Se3_hHxNi_HSE.out

_cell_length_a                         7.611613
_cell_length_b                         7.612295
_cell_length_c                         8.268893
_cell_angle_alpha                      114.286545
_cell_angle_beta                       65.716757
_cell_angle_gamma                      151.764065
_symmetry_space_group_name_H-M         'P 1'
_symmetry_Int_Tables_number            1

loop_
_symmetry_equiv_pos_as_xyz
   'x, y, z'

loop_
   _atom_site_label
   _atom_site_type_symbol
   _atom_site_fract_x
   _atom_site_fract_y
   _atom_site_fract_z
Ti001  Ti  -0.027058   0.027248   0.029691
Ti002  Ti  -0.344392   0.344435  -0.167684
 O003   O  -0.481447   0.481513   0.028351
 O004   O   0.129831  -0.129784  -0.104054
 O005   O  -0.036103   0.036090  -0.166333
 O006   O  -0.217258   0.217172  -0.233351
Pb007  Pb   0.401345  -0.400551  -0.265584
Pb008  Pb   0.286188  -0.286750   0.481848
Pb009  Pb  -0.373827   0.373253   0.194281
Se010  Se   0.313328  -0.312174   0.137450
Se011  Se  -0.369334   0.368208  -0.446803
Se012  Se  -0.021975   0.022040   0.373089

# end of cif
\end{verbatim}

\end{document}